# Monte Carlo Simulation for Polychromatic X-ray Fluorescence Computed Tomography with Sheet-Beam Geometry

Shanghai Jiang [1], Peng He [1, 2*], Luzhen Deng[1, 3], Mianyi Chen[1], Biao Wei [1, 2*]

*1 Key Lab of Optoelectronic Technology and Systems, Ministry of Education, Chongqing University, Chongqing, 400044, China*
*2. Engineering Research Center of Industrial Computed Tomography Nondestructive Testing, Ministry of Education, Chongqing University, Chongqing 400044, China*
*3. Department of Radiation Physics, the University of Texas MD Anderson Cancer Center, Houston, TX 77030, USA*

* Corresponding author. Peng He, penghe@cqu.edu.cn; Biao Wei, weibiao@cqu.edu.cn

**Abstract:** X-ray fluorescence computed tomography based on sheet-beam can save a huge amount of time to obtain a whole set of projections using synchrotron. However, it is clearly unpractical for most biomedical research laboratories. In this paper, polychromatic X-ray fluorescence computed tomography with sheet-beam geometry is tested by Monte Carlo simulation. First, two phantoms (A and B) filled with PMMA are used to simulate imaging process through GEANT 4. Phantom A contains several GNP-loaded regions with the same size (10 mm) in height and diameter but different Au weight concentration ranging from 0.3% to 1.8%. Phantom B contains twelve GNP-loaded regions with the same Au weight concentration (1.6%) but different diameter ranging from 1mm to 9mm. Second, discretized presentation of imaging model is established to reconstruct more accurate XFCT images. Third, XFCT images of phantom A and B are reconstructed by fliter backprojection (FBP) and maximum likelihood expectation maximization (MLEM) with and without correction, respectively. Contrast to noise ratio (CNR) is calculated to evaluate all the reconstructed images. Our results show that it is feasible for sheet-beam XFCT system based on polychromatic X-ray source and the discretized imaging model can be used to reconstruct more accurate images.

## 1. INTRODUCTION

As a promising imaging modality, X-ray computed tomography (XFCT) combined x-ray analysis and tomographic reconstruction algorithm has attracted wide concern in recent years. It can not only measure the distribution of elements but also the content of elements within samples in a nondestructive and noninvasive manner[1-3]. Conventional XFCT techniques with synchrotron source scan samples using the translation-rotation method, which is obviously unsuitable for most biomedical research laboratories due to its huge and expensive equipment.

Some improvements were proposed by other researchers. The synchrotron source is replaced by X-ray tube, simulated and experimental demonstration of polychromatic-source XFCT were implemented to reduce dose and scan time, which benchtop system feasible[4-6]. Although XFCT based on sheet beam geometry using synchrotron were also developed, fewer researches were done with polychromatic X-ray source[7].

As a contrast agent, Gold Nanoparticles (GNPs) have attracted wide concern due to its application in cancer detection and therapy [8-10]. Au within samples exposed by X-ray beam mainly emits K-shell X-rays fluorescence, which can be detected to reconstruct XFCT images. In this study, sheet-beam XFCT system based on polychromatic X-ray source was verified by Monte Carlo simulation. First, two phantoms contained several GNP-loaded regions were imaged through Geant4. The XFCT images were reconstructed by FBP and MLEM algorithms. Second, discretized presentation of sheet-beam geometry was established to reconstruct more accurate XFCT images. Contrast-to-noise ratio (CNR) was used to evaluate images quality. At last, CNR for the reconstructed images as functions of Au weight concentration and size of GNP-loaded regions was discussed.

## 2. Principles and methods

Principles and simulations of polychromatic X-ray Fluorescence computed tomography with sheet-beam geometry are presented in this paper. The data analysis techniques and image reconstruction algorithm are described in Secs. 2. 1-2. 4.

### 2. 1. Image system

The schematic diagram of Sheet-Beam CT system proposed in our study is shown in Fig.1. The system includes polychromatic sheet beam x-ray source, parallel collimator, array detectors, spectrometer, and computer.

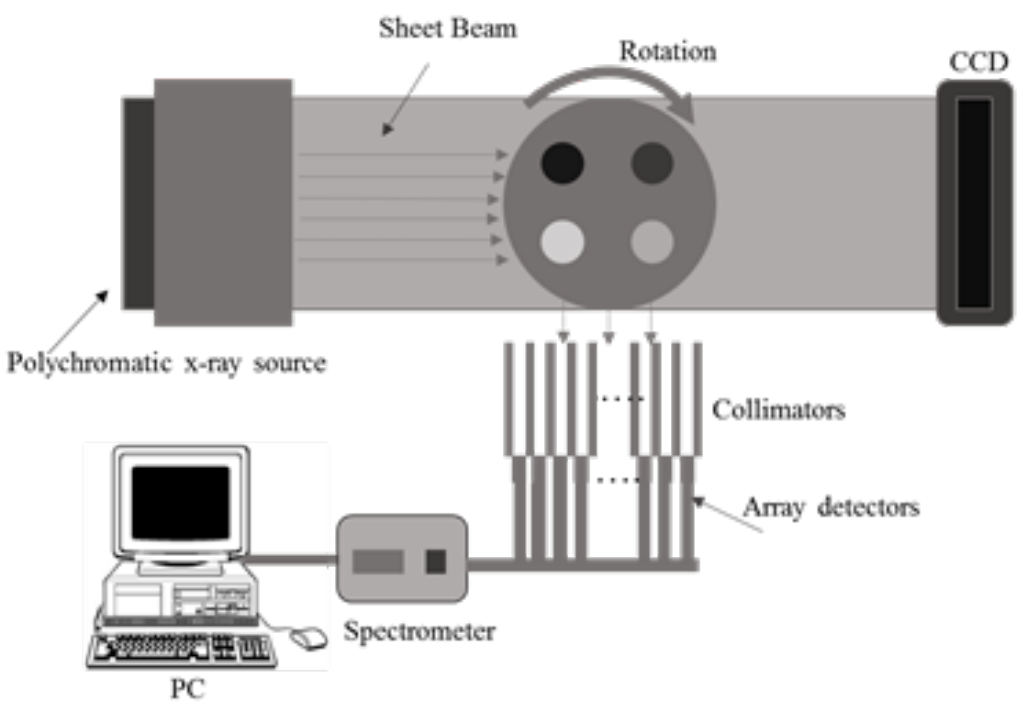

Fig.1. Schematic diagram of XFCT imaging system using sheet-beam and linear detector arrays.

Polychromatic X-rays from X-ray tube is collimated into sheet beam, and then impinges on the object to cover the whole cross-section. GNPs exposed by X-ray beam can isotropically emit characteristic X-ray photons. Linear array photon-counting detectors with energy resolution are positioned perpendicular to the beam propagation direction for x-ray fluorescent spectra[11].

#### 2. 1.1. Monte Carlo model

The Monte Carlo simulation was implemented by GEANT 4 software. Because simulations of projection at each angle are independent of each other, we replace the X-ray source in Fig.1 with a virtual source to reduce simulation time of GEANT 4. The spectrum of sheet-beam X-ray source was calculated by SpekCal software, which simulates X-ray spectra emitted from thick-target tungsten anode X-ray tubes [10, 12]. In this study, electron beam of 120 keV interacted with tungsten. Then, the emitted X-ray photons were filtered by Sn with thickness of 1 mm. The spectrum of X-ray source was shown in Fig.2. Here, the width and thickness of sheet beam were set to 6.4*cm* and 1*mm*, respectively.

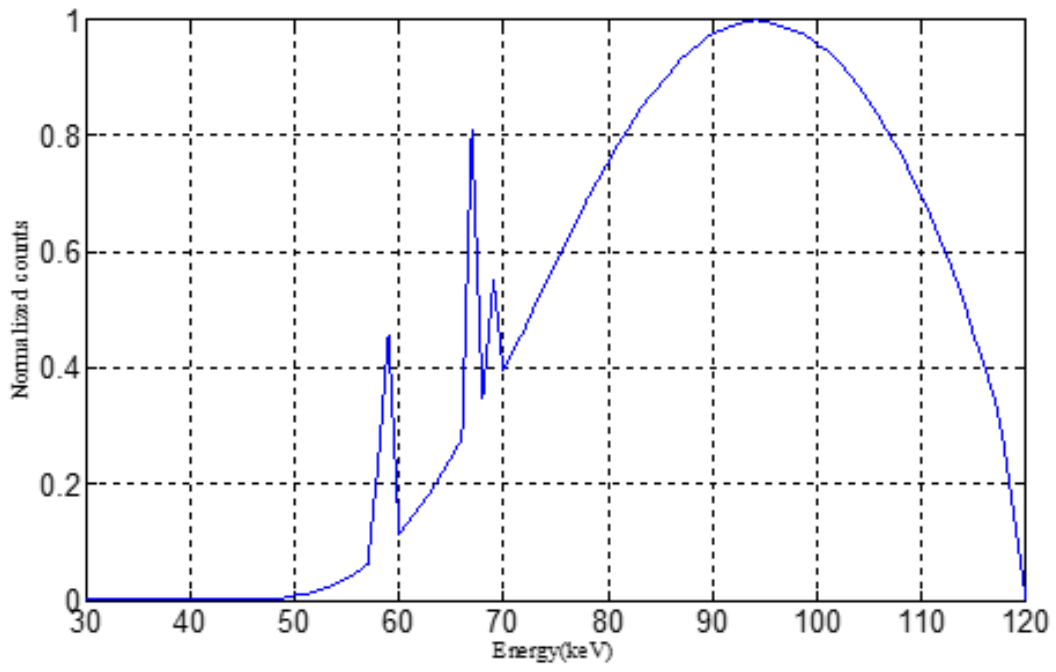

Fig.2. Spectrum of incident polychromatic X-ray source.

Two GNP-loaded PMMA phantoms are shown in Fig.3. The PMMA phantom is 6.4cm in both height and diameter. The phantom on the left side contained several GNP-loaded regions, which has the same size (10mm) in height and diameter but different gold concentration (mixed with water) ranging from 0.3% to 1.8%. The GNP-loaded regions in right phantom has the same concentration (1.5%) but different diameter ranging from 1mm to 9mm.

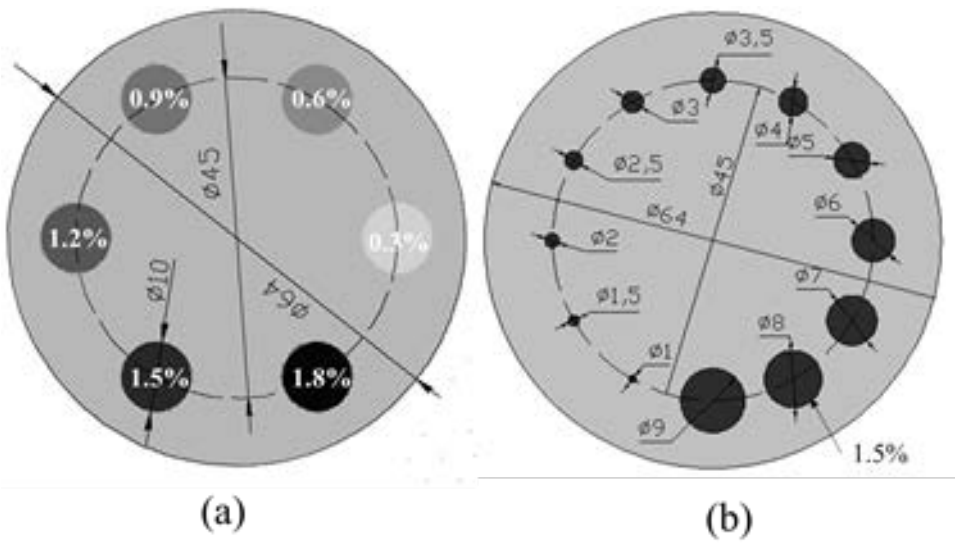

Fig.3. Phantoms contained GNP-loaded regions. (a) GNP-loaded regions with same size in height and diameter but different Au weight concentrations; (b) GNP-loaded regions with same Au weight concentration but different diameters.

The emitted fluorescence photons were detected by a series of energy-sensitive tallies (shown in Fig.1.). They were positioned 1mm behind lead collimator with a series of pinhole openings of diameter 0.5mm. The whole XFCT scanning procedure was divided into independent simulation for each projection angle. The deterministic point detector tally (F5) and E card were used to simulate 64 energy sensitive detectors, where each detector has the same sensitive area (0.5mm × 0.5mm) and energy resolution (0.5keV) [13].When 10M histories (photons) were calculated for each simulation, the uncertainty was less than 5% for relevant photon energies (50-75keV).

### 2.1.2. Data acquisition

The X-ray photons arriving at the detectors mainly comes from both Compton scatter and characteristic x-ray photons. Considering fluorescent field and the attenuation of low-energy photon in the phantom, the gold $K_\alpha$ lines (67.0 and 68.8keV) are the best candidates to reconstruct XFCT images in our simulations. To extract fluorescent signal count, cubic polynomial was used to fit the points around the gold fluorescent peaks. The fluorescence signal counts measured of each projection during the simulation was the difference between the measured signal counts and the fitted counts[10, 14]. A sinogram of the gold fluorescence signal counts was reconstructed using the extracted gold fluorescence signal from each projected simulation.

## 2.2. The imaging model of sheet-beam XFCT

The imaging model of sheet-beam XFCT were established previously by some researches[7, 11] , and its geometry is presented in Fig.4. While the *xy*-coordinate system is attached to an object, the *st*-coordinate system is spun with the data acquisition system, and can be at any instant obtained by rotating the *xy*-coordinate system by an angle θ counterclockwise[15]. That is, their relationship can be expressed as follows:

$$\begin{pmatrix} s \\ t \end{pmatrix} = \begin{pmatrix} \cos\theta & \sin\theta \\ -\sin\theta & \cos\theta \end{pmatrix} \begin{pmatrix} x \\ y \end{pmatrix} \tag{1}$$

According to the results of previous research [11, 16-18], the total photons of fluorescent x-ray reaching the *i*th detector is represented as follows:

$$I_{iv}(\theta, s) = \int_{-\infty}^{+\infty} f(\alpha, s, t) \bullet g(\alpha, s, t) \bullet \Omega(s, t)\, \rho(x, y) dt \tag{2}$$

Where

$$f(\theta, s, t) = I_0 \exp[-\int_{-\infty}^{t} \mu^I(x, y) ds] \tag{3}$$

$$g(\theta, s, t) = \mu_{ph} \omega \int_{\gamma_{\min}}^{\gamma_{\max}} \exp[-\int_0^{\infty} \mu^F(x, y) db] d\gamma \tag{4}$$

$\omega$ is the yield of characteristic X-ray photons. $\Omega$ is the solid angle at which the point, *Q*, is viewed by the by the *m*th fluorescence detector. $\mu_{ph}$ is the photoelectric linear attenuation coefficient of Au. The $\rho(x, y)$, $\mu^I(x, y)$, $\mu^F(x, y)$ are the distribution of Au weight concentration, linear attenuation coefficient of incident x-ray energy and linear coefficient of fluorescent x-ray. Here, in order to simplify reconstruction, the equation (4) can be expressed approximately as follows:

$$I_{iv}(\theta, s) \approx \mu_{ph} \omega I_0 \Omega \int_{-\infty}^{+\infty} \rho(x, y) dt \tag{5}$$

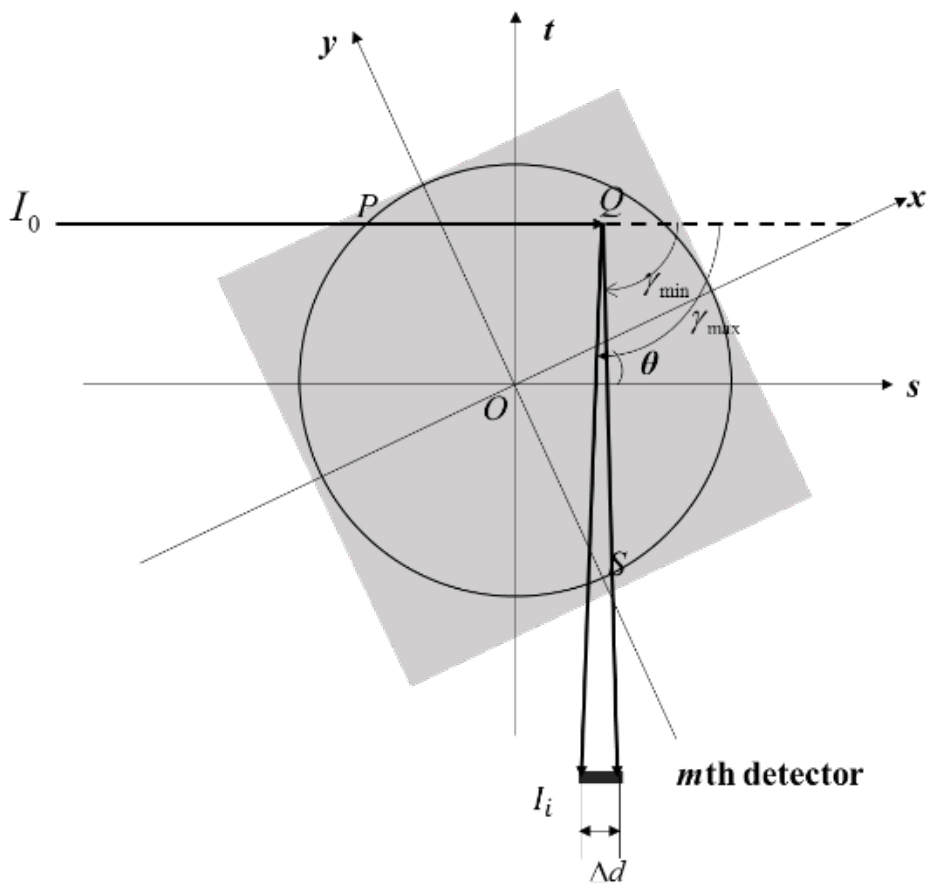

Fig.4. Schematic diagram of XFCT imaging geometry using sheet-beam.

Thus, the measured process by the XFCT based on sheet beam geometry can be viewed approximately as Radon transform, and FBP algorithm can be used to reconstruct XFCT images. Here, the reconstructed images are usually considered to be uncorrected.

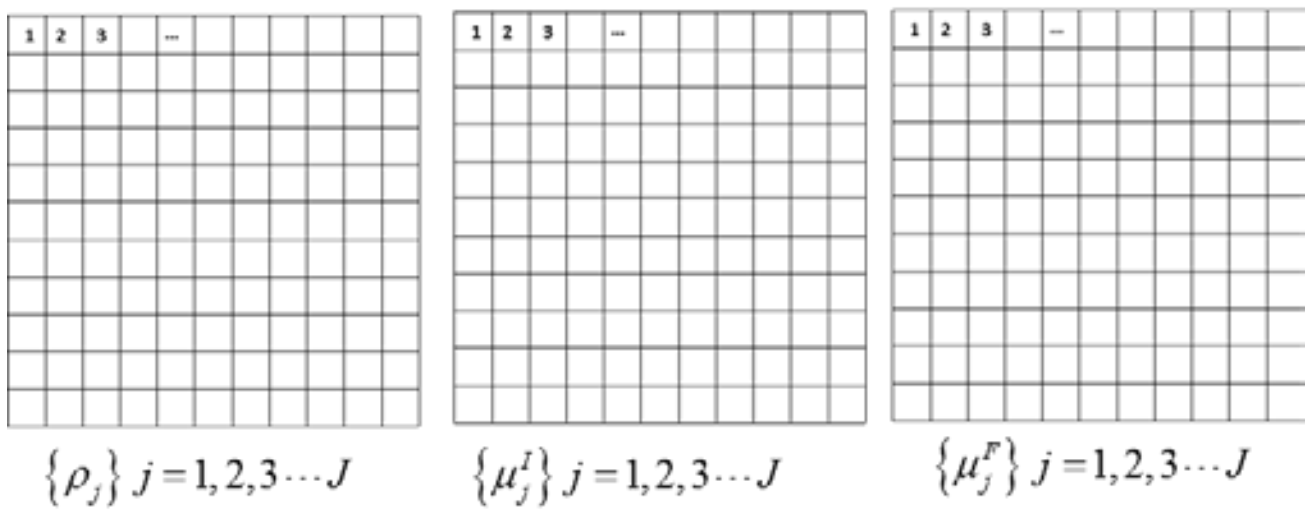

Fig.5. The matrices defined for XFCT

To acquire the corrected images, it is necessary to obtain discretized representation of equation (2). During the process, we assume that the phantom is two–dimensional. Three matrices shown in Fig.5, including $\rho_j$ , $\mu_j^I$ and $\mu_j^F$ corresponding to $\rho(x,y)$ , $\mu^I(x,y)$ and $\mu^F(x,y)$, are used to describe the whole phantom, where *j* ( $j=1,2,3...J$ ) represents the number of each pixel. We assume that sheet-beam X-ray incident phantom at *N* angles, where *n* ( $n=1,2,3\dots N$ ) represents the number of each angle. The sheet-beam source is considered as *P* X-rays at each incident direction shown in Fig 6(a). Here, $p'$ is the number of the $p'$ th X-ray. We consider the process of the *p*th incident X-ray interacted with phantom. Therefore,

$$p = P(n-1) + p' \tag{6}$$

*Step 1*: Let $Q_{pj}$ the subset of $Q_p$, which consists of the light blue pixels shown in Fig.6(b). $L_{pj}$ is the intersected length of the *p*th X-ray and the *j*th pixel. The incident X-ray intensity before reaching the *j*th pixel is expressed as follows:

$$f_{pj} = I_0 \exp(-\sum_{k\in Q_{pj}} \mu_k^I L_{pk}^I) \tag{7}$$

*Step 2:* The total X-ray counts emitted from the *j*th pixel is proportional to the product of $\rho_j$ and $\omega\mu_{ph} f_{pj} L_{pj}^I$ . Let $\delta_{ij}$ is the angle viewed by the detector corresponding to *i*th projection at the *j*th pixel. Here, we assume that the X-ray counts emitted from the *j*th pixel can be recorded by the *m*th detector, where number *i* can be calculated by equation (6). The X-ray counts measured by the detector are written as follows:

$$\mu_{ph}\omega\delta_{ij} f_{pj}\rho_j L_{pj}^I \tag{8}$$

*Step 3:* Not all X-ray fluorescence photons emitted from *j*th pixel can be detected by detectors. Here, we assume that the X-ray photos by the *j*th pixel can be detected within $\delta_{ij}$ when the line passing through the center of the *j*th pixel and paralleling to the lead hole can reach detector without block. Attenuation of fluorescent X-ray from *j*th to detector must be also considered during the further process. In Fig.6(c), the fan-shaped X-ray can be divided into *K* (*K* is positive integer) individual X-rays. Let $\Delta\delta = \delta_{pj} / K$ and *l* ( $1 \le l \le K$ ) is the number of individual fluorescent x-rays. The attenuation of the *l*th fluorescent X-ray can be expressed as follows:

$$\exp\left(-\sum_{q\in T_{pjl}} \mu_q^F L_{pjq}^F\right) \tag{9}$$

Where $T_{pjl}$ is defined as the set consisting of the pixels (light blue squares shown in Fig.6(d)) intersected with the $l$th fluorescent X-ray. $L_{pjq}^{F}$ is described as the intersected length of the $l$th fluorescent X-ray with the $q$th pixel ( $q \in T_{pjl}$ ).

$$g_{pij} = \mu_{ph} \omega \delta_{pj} \sum_{l=1}^{K} \exp\left( - \sum_{q \in T_{pjl}} \mu_q^F L_{pjq}^F \right) \tag{10}$$

*Step4:* We consider the process of the $p$th incident X-ray interacted with phantom. Let us define that $i$ is the number of $i$th projection ranging from 1 to $I$. The contribution $h_{ij}$ of the $j$th pixel to the $i$th projection at $p$th incident X-ray can be expressed as follows:

$$i = M(n-1) + m \tag{11}$$

$$h_{ij} = f_{pj} g_{pij} \delta_{ij} \tag{12}$$

Accordingly, the discretized representation of (4) is written as follows:

$$I_i = \sum_j h_{ij} \rho_j \quad (i = 1, 2, \ldots, I) \tag{13}$$

The matrix representation of (9) is

$$\boldsymbol{I} = \boldsymbol{H} \boldsymbol{\rho} \tag{14}$$

Where

$$\boldsymbol{H} = (h_{ij}) \quad (1 \le i \le M, 1 \le j \le N) \tag{15}$$

$$\boldsymbol{I} = (I_i) \quad (1 \le i \le M) \tag{16}$$

and

$$\boldsymbol{\rho} = (\rho_j) \quad (1 \le j \le N) \tag{17}$$

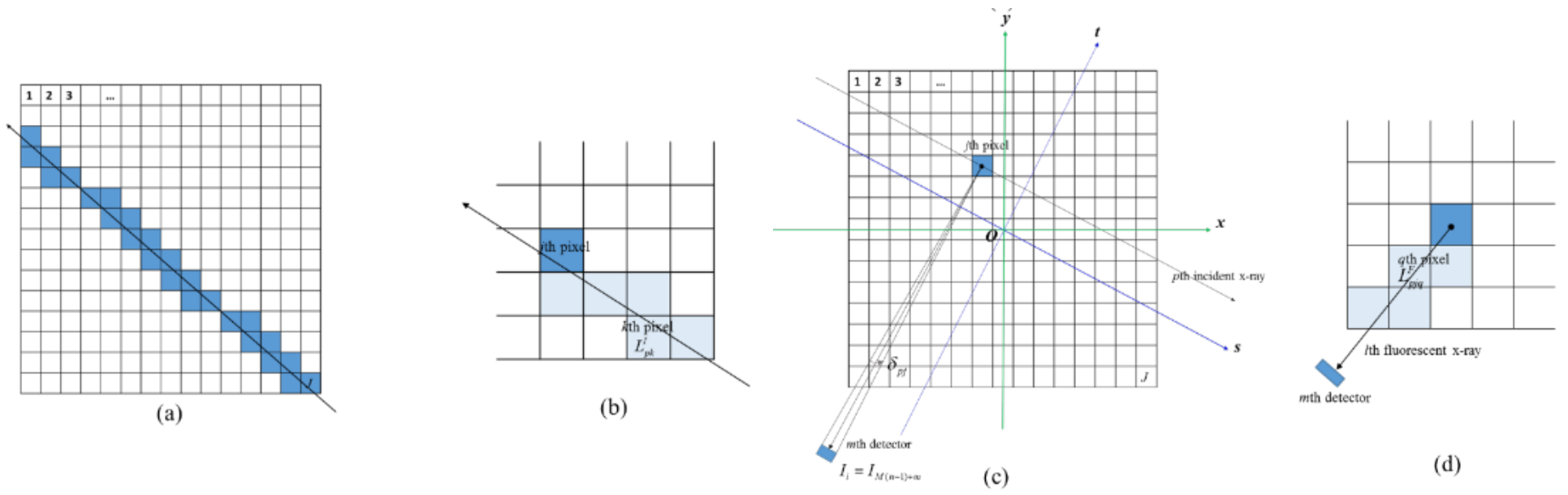

Fig.6. The parameters defined for discretized presentation. (a) a set $Q_p$ defined for blue squares intersected by the $p$th incident X-ray.(b) a set $Q_{pj}$ defined for light blue squares intersected by the $p$th incident X-ray. (c) definition of $\delta_{ij}$, $m$. (d) a set $T_{pjq}$ defined for light blue squares intersected by $l$th fluorescent X-ray.

## 2.3. XFCT image reconstruction

Here, we assume that the maps of $\mu^I$ and $\mu^F$ are known in our study. Two algorithms, including FBP and MELM, were used to reconstruct XFCT images with correction and without correction, respectively. First, sinograms of two phantoms with GNP-loaded regions were acquired by the described method in section 2.1.2. Then, XFCT images with $64\times64$ pixels of $64\times64$ $mm^2$ were reconstructed by FBP and MLEM without correction. The more accurate projection matrix described in equation (6)-(18) were calculated to correct the effect of attenuation for high quality images.

### 2.4. XFCT image analysis

The reconstructed XFCT images are evaluated as CNR by calculating the ratio of difference between the mean value of each GNP-loaded region and background (PMMA) and standard deviation of background. CNR is defined as follows [15, 19]:

$$CNR = \frac{\overline{\Psi}_{\mathrm{Re}gion} - \overline{\Psi}_{BK}}{V_{BK}} \tag{18}$$

Where $\overline{\Psi}_{\mathrm{Re}gion}$ and $\overline{\Psi}_{BK}$ are mean reconstructed values of GNP-loaded region and background, $V_{BK}$ is standard deviation of background (PMMA). According to the Rose criterion, imaging sensitivity limit of the system proposed was determined using CNR of 4[20].

## 3. RESULTS

### 3.1. XFCT image reconstruction

Sinograms of two phantoms with GNP-loaded regions are shown in Fig.7. Fig.7(a) shows the Sinogram of phantom *A* and Fig.7(b) is the sinogram of phantom *B*. The reconstructed XFCT images of phantom A are shown in Fig.8, where Fig.8(a) and Fig.8(c) are the images reconstructed by FBP and MLEM (100 iterations) without correction, respectively. Correspondingly, Fig.9(b) and Fig.9(d) are reconstructed by FBP and MLEM (100 iterations) with correction, respectively. Similar representation is also shown in Fig.9. Obviously, grey values of GNP-loaded regions decrease with reduction of Au weight concentration shown in Fig.8.

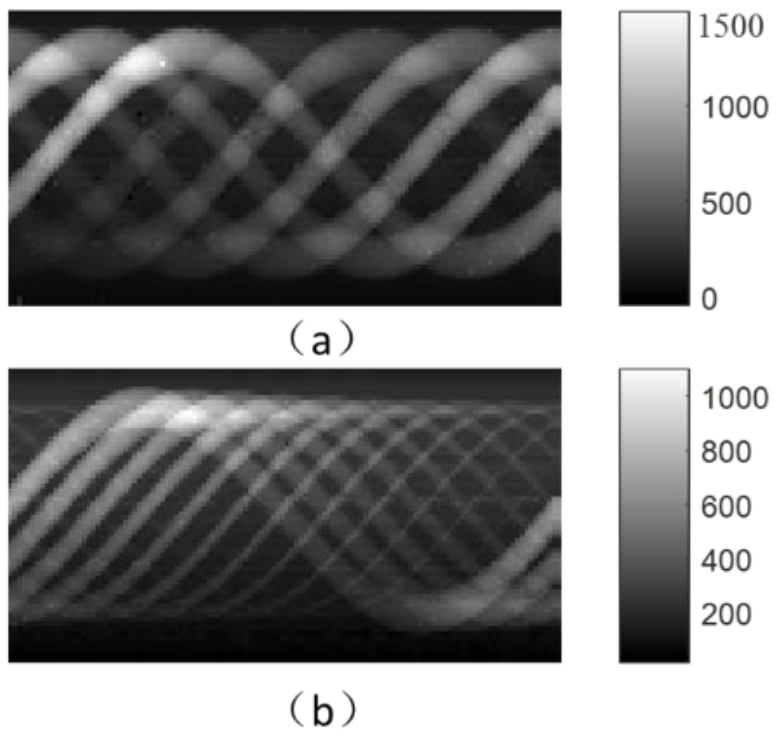

Fig.7. The reconstructed sinograms of (a) phantom *A* and (b) phantom *B*.

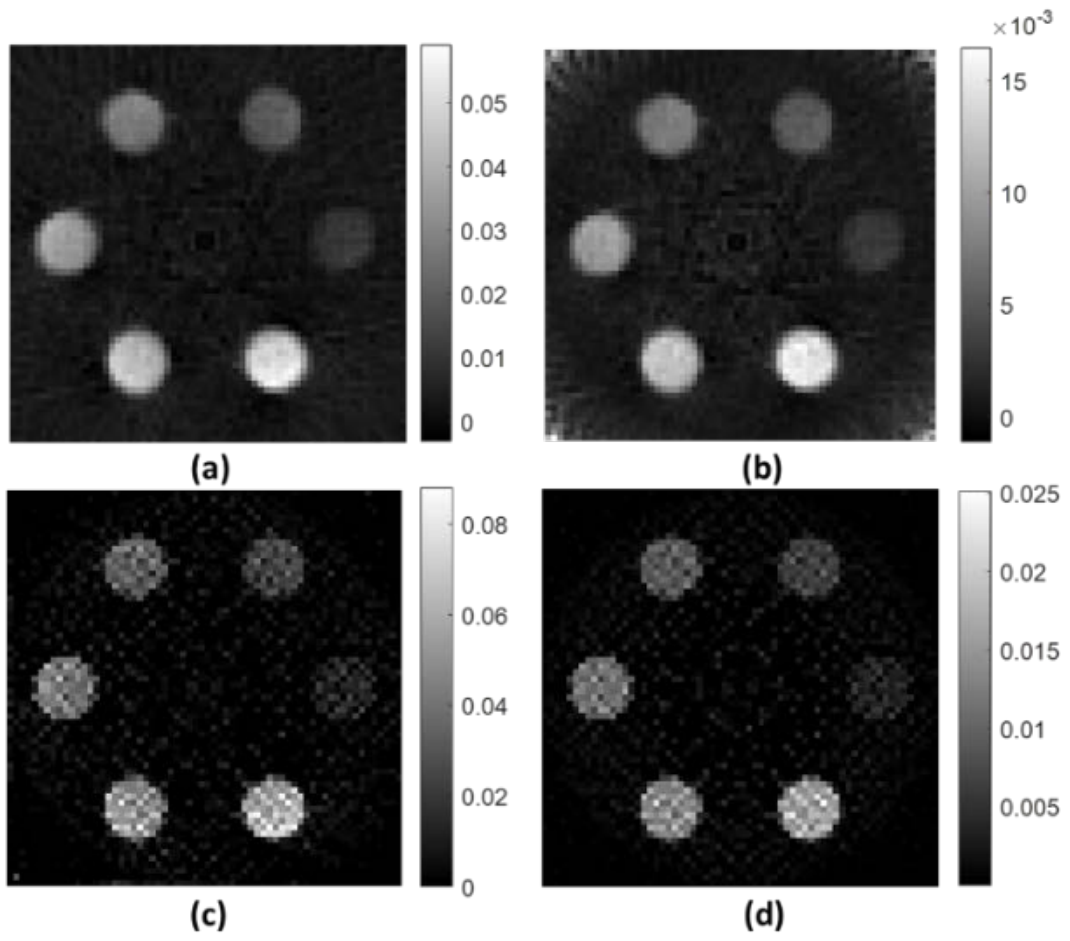

Fig.8. Reconstructed XFCT images of phantom *A*. (a) reconstructed by FBP without correction, (b) reconstructed by FBP with correction, (c) reconstructed by MLEM without correction and (d) reconstructed by MLEM with correction.

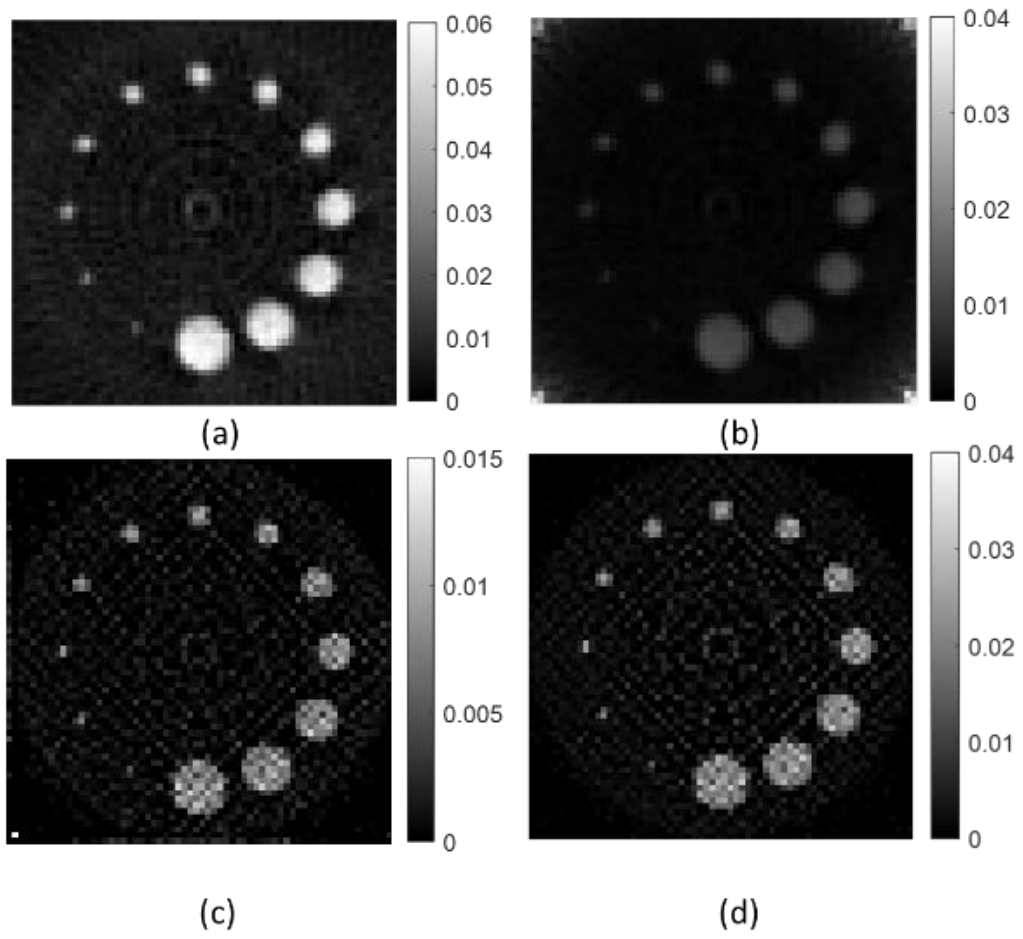

Fig.9. Reconstructed XFCT images of phantom *B*. (a) reconstructed by FBP without correction, (b) reconstructed by FBP with correction, (c) reconstructed by MLEM without correction and (d) reconstructed by MLEM with correction.

The reconstructed Au concentration calculated from the mean value of each GNP-loaded region in phantom A is plotted in Fig.10(a) (acquired by FBP) and Fig.10(b) (acquired by MLEM). Both figures show that uncorrected Au weight concentration has larger error than corrected concentration, which may mean that our corrected model can provide more accurate results. For the same Au weight concentration, the reconstructed values in Fig. 10(d) acquired by MLEM algorithm has more stable values than in Fig.10(c) acquired by FBP algorithm.

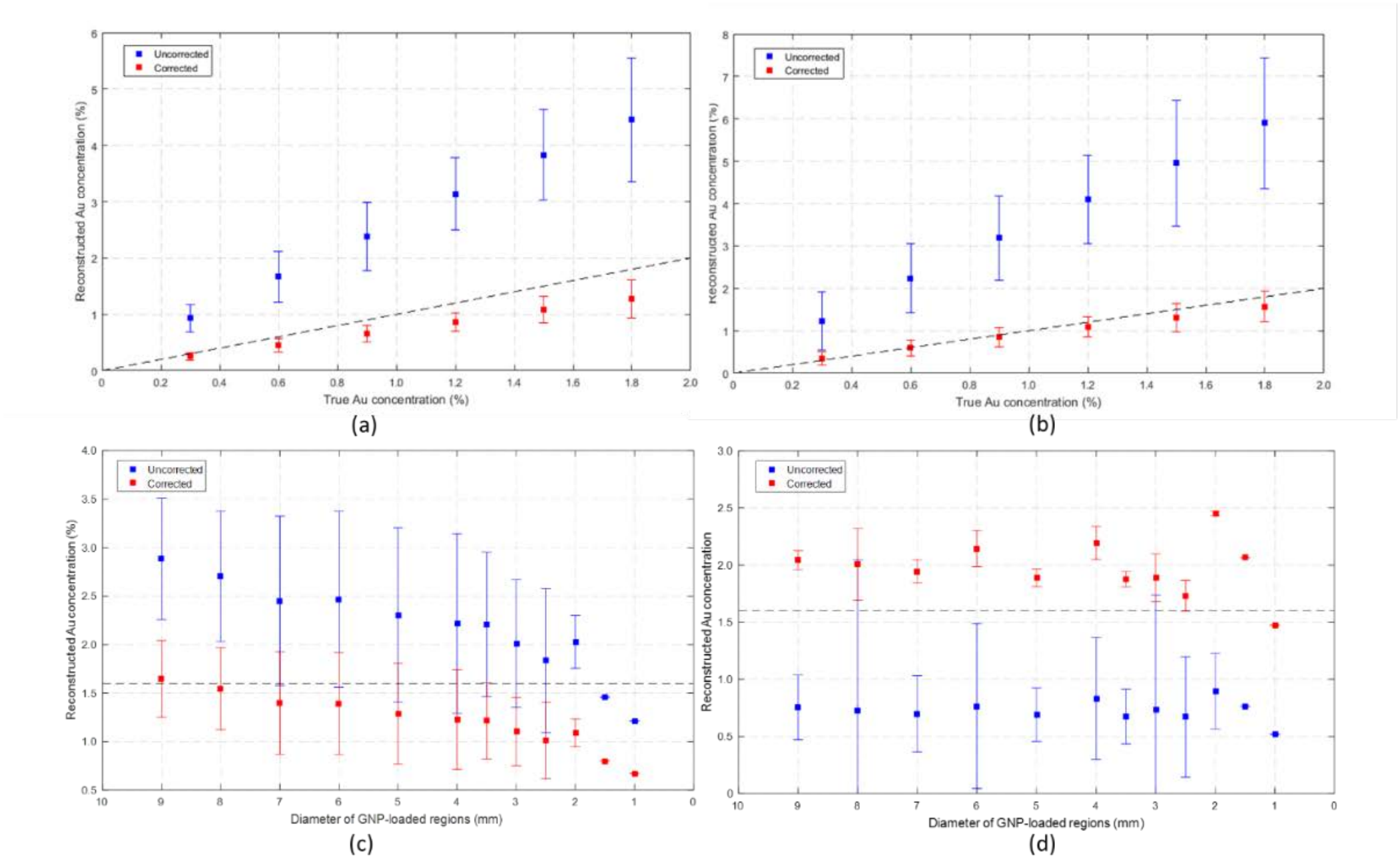

Fig.10. Reconstructed Au weight concentration. (a) and (c) acquired by FBP algorithm with and without correction. (b) and (d) acquired by MLEM algorithm with and without correction.

### 3.2. XFCT image analysis

CNR for reconstructed XFCT images with FBP and MLEM as a function of Au weight concentration is presented in Fig.11(a) and Fig.11(b), respectively. Both of bar charts show us that values of CNR in our setup has higher than 4, when Au weight concentration is greater than 0.6%. To detect lower concentration, the setup needs be modified, including length and diameter of collimators, spectrum of x-ray source, distance from x-ray source to the center of phantom, and so on[21]. Fig.11 (a) and Fig.11 (b) can also indicate that CNR of GNP-loaded region for same size is linear proportional to Au weight concentration ( $R^2 \geq 0.9992$ ). According to Rose criterion (CNR>4), not all GNP-loaded regions in Fig.8 were detectable, and detection limits from Fig.8(a) to Fig.8(d) were 0.59%, 0.62%, 0.60% and 0.56%, respectively.

CNR for reconstructed XFCT images with FBP and MLEM as a function of each GNP-loaded region size is also presented in Fig.11(c) and Fig. 11(d), respectively. When the diameter of GNP-loaded region is ranging from 7mm to 9mm, the values of CNR increase with increase of diameter but fluctuate largely from 1mm to 6mm. The phenomenon illustrates that size of GNP-loaded region can influence values of CNR when incident X-ray width is invariant.

Algorithms can influence values of CNR. In Fig.11(a) and Fig.11(c), the corrected images have lower CNR than uncorrected ones with FBP. On the contrary, the corrected images with MLEM have greater than the uncorrected ones in Fig.11(b) and Fig.11(d). However, the corrected concentrations have more accurate than the uncorrected ones.

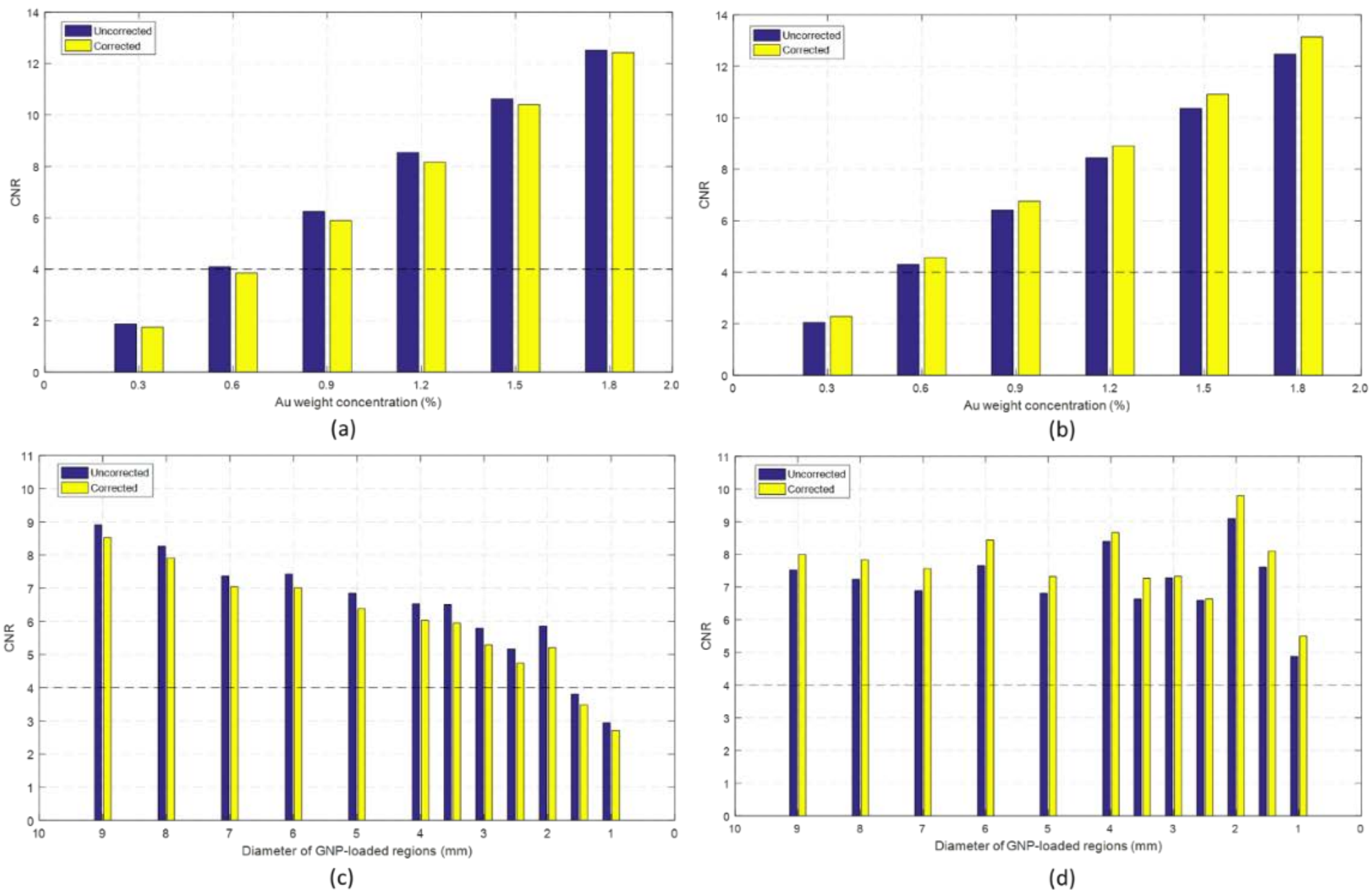

Fig.11. CNR. (a) and (c) acquired by FBP algorithm with and without correction . (b) and (d) acquired by MLEM algorithm with and without correction.

## 4. Discussion

We have presented a benchtop system for polychromatic X-ray fluorescence computed tomography with sheet-beam geometry through Monte Carlo simulation. The discretized model with sheet-beam XFCT are also described in our study.

In the simulation, we used polychromatic X-rays (X-ray tubes) instead of synchrotron radiation in similar imaging system proposed previously by others,[7] which make it possible to reduce costs and size of apparatus. Another advantage of XFCT imaging with sheet-beam geometry is a drastic reduction of overall scanning time, compared to traditional XFCT[10]. Although the proposed XFCT system may not be demonstrated by experimental study, it may provide valuable method for optimization of XFCT system.

However, a technical challenge for the proposed system is improvement of detection limit and CNR. First, they may be improved further by additional modifications to the current setup such as quasi-monochromatization of incident x-ray spectrum and further optimization of detector collimation[7, 18]. Second, the optimized algorithm may improve image quality. According to our results, different algorithms influence the values of CNR, which is similar with conclusion of Zichao.et.al.[22] Thirdly, X-ray detector with higher energy resolution is used during the process. Another potential approach is to consider both $K_\alpha$ peaks and $K_\beta$ peaks at the same time.

Our future work will consist of optimizing polychromatic XFCT with sheet-beam geometry, such as spectrum of X-ray source and length of collimators. We will also build an imaging system based on our simulations to demonstrate the feasibility by experiment.

## 5. Conclusion

In this investigation, the feasibility of polychromatic sheet-beam XFCT system proposed in this study was demonstrated by Monte Carlo method. Two phantoms contained several GNP-loaded regions were imaged using GEANT 4. Accurate images were reconstructed by FBP and MLEM with and without correction , respectively. Our results may provide necessary justification for the design of benchtop XFCT imaging system for in vivo imaging.

### Disclosure of conflict of interest

The authors declare that there is no conflict of interest regarding the publication of this paper.

## Acknowledgements

This work is partially supported by the National Natural Science Foundation of China (61401049), the Graduate Scientific Research and Innovation Foundation of Chongqing, China (Grant No. CYB16044), the Fundamental Research Funds for the Central Universities (106112016CDJXY120003), Chongqing strategic industry key generic technology innovation project (NO. cstc2015zdcy-ztzxX0002).

## References

1. G. R. Pereira, R. T. Lopes, M. J. Anjos, H. S. Rocha and C. A. Pérez, "X-ray fluorescence microtomography analyzing reference samples," *Nucl. Instr. Meth. Phys. Res.*, vol. 579, no. 1, pp. 322-325, 2007.
2. D. H. Mcnear, P. Edward, E. Jeff, R. L. Chaney, S. Steve, N. Matt, R. Mark and D. L. Sparks, "Application of quantitative fluorescence and absorption-edge computed microtomography to image metal compartmentalization in Alyssum murale," *Environ. Sci. Technol.*, vol. 39, no. 7, pp. 2210-2218, 2005.
3. T. Takeda, Q. Yu, T. Yashiro, T. Zeniya, J. Wu, Y. Hasegawa, Thet-Thet-Lwin, K. Hyodo, T. Yuasa and F. A. Dilmanian, "Iodine imaging in thyroid by fluorescent X-ray CT with 0.05 mm spatial resolution," *Nucl. Instr. Meth. Phys. Res.*, vol. 467–468, no. 7, pp. 1318–1321, 2001.
4. K. Ricketts, C. Guazzoni, A. Castoldi and G. Royle, "A bench-top K X-ray fluorescence system for quantitative measurement of gold nanoparticles for biological sample diagnostics," *Nuclear Instruments and Methods in Physics Research Section A: Accelerators, Spectrometers, Detectors and Associated Equipment*, vol. 816, pp. 25-32, 2016.
5. N. Manohar and S. H. Cho, "Quality of micro-CT images acquired from simultaneous micro-CT and benchtop x-ray fluorescence computed tomography (XFCT): A preliminary Monte Carlo study," in *Nuclear Science Symposium and Medical Imaging Conference (NSS/MIC), 2013 IEEE*, Ed., pp. 1-3, IEEE, 2013.
6. Y. Kuang, G. Pratx, M. Bazalova, B. Meng, J. Qian and L. Xing, "First Demonstration of Multiplexed X-Ray Fluorescence Computed Tomography (XFCT) Imaging.pdf>," *Medical Imaging, IEEE Transactions on*, vol. 32, no. 2, pp. 262-267, 2013.
7. Q. Huo, T. Yuasa, T. Akatsuka, T. Takeda, J. Wu, Thet-Thet-Lwin, K. Hyodo and F. A. Dilmanian, "Sheet-beam geometry for in vivo fluorescent x-ray computed tomography: proof-of-concept experiment in molecular imaging," *Opt. Lett.*, vol. 33, no. 21, pp. 2494, 2008.
8. M. J. Pushie, I. J. Pickering, M. Korbas, M. J. Hackett and G. N. George, "Elemental and chemically specific X-ray fluorescence imaging of biological systems," *Chem. Rev.*, vol. 114, no. 17, pp. 8499-8541, 2014.
9. L. E. Cole, R. D. Ross, J. M. Tilley, T. Vargo-Gogola and R. K. Roeder, "Gold nanoparticles as contrast agents in x-ray imaging and computed tomography," *Nanomedicine*, vol. 10, no. 2, pp. 321-341, 2015.
10. B. L. Jones and S. H. Cho, "The feasibility of polychromatic cone-beam x-ray fluorescence computed tomography (XFCT) imaging of gold nanoparticle-loaded objects: a Monte Carlo study," *Phys. Med. Biol.*, vol. 56, no. 12, pp. 3719-3730, 2011.
11. S. Nakamura, Q. Huo and T. Yuasa, "Reconstruction technique of fluorescent x-ray computed

tomography using sheet beam," in *Signal Processing Conference (EUSIPCO), 2014 Proceedings of the 22nd European*, Ed., pp. 1975 - 1979, 2014.

12. G. Poludniowski, G. Landry, F. DeBlois, P. M. Evans and F. Verhaegen, "SpekCalc: a program to calculate photon spectra from tungsten anode x-ray tubes," *Phys. Med. Biol.*, vol. 54, no. 19, pp. N433-438, 2009.
13. N. Sunaguchi, T. Yuasa, K. Hyodo and T. Zeniya, "The feasibility study on 3-dimensional fluorescent x-ray computed tomography using the pinhole effect for biomedical applications," in *Engineering in Medicine and Biology Society (EMBC), 2013 35th Annual International Conference of the IEEE*, Ed., pp. 2348-2351, IEEE, 2013.
14. S. K. Cheong, B. L. Jones, A. K. Siddiqi, F. Liu, N. Manohar and S. H. Cho, "X-ray fluorescence computed tomography (XFCT) imaging of gold nanoparticle-loaded objects using 110 kVp x-rays," *Phys. Med. Biol.*, vol. 55, no. 3, pp. 647-662, 2010.
15. P. Feng, W. Cong, B. Wei and G. Wang, "Analytic Comparison between X-ray Fluorescence CT and K-edge CT," *IEEE Trans. Biomed. Eng.*, vol. 61, no. 3, pp. 975-985, 2014.
16. Q. Yang, B. Deng, W. Lv, F. Shen, R. Chen, Y. Wang, G. Du, F. Yan, T. Xiao and H. Xu, "Fast and accurate X-ray fluorescence computed tomography imaging with the ordered-subsets expectation maximization algorithm," *J.Synchrotron Radiat.*, vol. 19, no. Pt 2, pp. 210-215, 2012.
17. P. J. Riviere, "Accelerating X-ray fluorescence computed tomography," *Conf. Proc. IEEE Eng. Med. Biol. Soc.*, vol. 2009, pp. 1000 - 1003, 2009.
18. Q. Huo, H. Sato, T. Yuasa, T. Akatsuka, J. Wu, T.-T. Lwin, T. Takeda and K. Hyodo, "First experimental result with fluorescent X-ray CT based on sheet-beam geometry," *X-Ray Spectrom.*, vol. 38, no. 5, pp. 439-445, 2009.
19. M. Bazalova-Carter, M. Ahmad, T. Matsuura, S. Takao, Y. Matsuo, R. Fahrig, H. Shirato, K. Umegaki and L. Xing, "Proton-induced x-ray fluorescence CT imaging," *Med. Phys.*, vol. 42, no. 2, pp. 900-907, 2015.
20. A. Rose, "Vision: Human and Electronic," *Phys. Today*, vol. 28, no. 12, pp. 49-50, 2008.
21. M. Bazalova, Y. Kuang, G. Pratx and L. Xing, "Investigation of x-ray fluorescence computed tomography (XFCT) and K-edge imaging," *Medical Imaging, IEEE Transactions on*, vol. 31, no. 8, pp. 1620-1627, 2012.
22. Z. Di, S. Leyffer and S. M. Wild, "Optimization-Based Approach for Joint X-Ray Fluorescence and Transmission Tomographic Inversion," *SIAM Journal on Imaging Sciences*, vol. 9, no. 1, pp. 1-23, 2016.